\documentclass[reprint,showpacs,prl,superscriptaddress]{revtex4-1}
\usepackage{epsfig}
\usepackage{amsmath}

\begin{document}

\title{Two-Photon Correlation of Broadband Amplified Spontaneous Four-Wave Mixing}
\author{Rafi Z. Vered, Michael Rosenbluh and Avi Pe'er}
\affiliation{Physics Department and BINA Center for Nano-technology, Bar-Ilan University, Ramat-Gan 52900, Israel}
\email{avi.peer@biu.ac.il}
\begin{abstract}
We measure the time-energy correlation of broadband, spontaneously seeded four wave mixing (FWM), and demonstrate novel time-frequency coupling effects; specifically, we observe a power-dependent splitting of the correlation in both energy and time. By pumping a photonic crystal fiber with narrowband picosecond pulses we generate FWM in a unique regime, where  broadband ($>100nm$), sidebands are generated that are incoherent, yet time-energy correlated. Although the observed time-energy correlation in FWM is conceptually similar to parametric down conversion, its unique dependence on pump intensity due to self and cross phase modulation effects, yields spectral and temporal structure in the correlations. While these effects are minute compared to the time duration and bandwidth of the FWM sidebands, they are well observed using sum frequency generation as a precise, ultrafast, wide-bandwidth correlation detector.
\end{abstract}
\maketitle

Four wave mixing (FWM), the nonlinear conversion of energy from a pair of pump frequencies $\omega_1,\omega_2$ to a pair of signal and idler frequencies $\omega_3,\omega_4$ such that $\omega_1\!+\!\omega_2\!=\!\omega_3\!+\!\omega_4$, is maybe the most prevalent nonlinear process in nature, which exists in practically any medium. Many applications rely on FWM for devices, such as ultra broadband amplifiers \cite{Hansryd2002applications} and widely tunable parametric oscillators \cite{Hasegawa1980tunable,Nakazawa1988Modulational}. In its time domain version of self phase modulation (SPM), FWM is responsible for the spectral broadening of short pulses to span over an octave of bandwidth, which enables stabilization of ultrafast frequency combs \cite{Holzwarth2000optical}.

Recently, frequency combs sources that are based on FWM oscillation in micro-cavities attracted attention \cite{okawachi2011octave, ferdous2011spectral,johnson2012chip,ferdous2012probing}. These sources are exciting for new avenues in comb based measurements and applications; yet, their temporal and spectral coherence properties are not fully understood. It is still unclear how and when broadband tooth-to-tooth coherence emerges during amplification of the spontaneous FWM seed, and whether FWM oscillators mode lock similar to standard lasers \cite{ferdous2011spectral}. Here, we directly measure the coherence properties of amplified spontaneous FWM in a simple, single-pass configuration, and demonstrate its unique two-photon coherence.

In terms of pump bandwidth, research in the past focused mainly on two extremes: either FWM pumped by a narrow CW pump (or relatively long $ns$ scale pulses), where FWM is purely spontaneously generated and no significant temporal effects exist; or, pumped by broadband $fs$ pulses, where the process is highly time dependent and dominated by stimulated self and cross phase modulation. The intermediate regime with picosecond pump pulses was far less explored. In this regime the pump is spectrally narrow and temporal effects are weak, leading to a unique combination of coherent spectral broadening in the vicinity of the pump due to self phase modulation (SPM), and incoherent generation of broadband sidebands far from the pump due to amplified spontaneous FWM \cite{vidne2005spatial}. This regime, the focus of this work, allows exploration of the nonlinear interplay between coherent and incoherent processes, and of the unique quantum and classical correlations associated with it.

While stimulated FWM is described very well by the classical equations of nonlinear optics, spontaneous FWM is inherently quantum, as it requires vacuum fluctuations to seed the conversion of a pair of pump photons into a pair of signal and idler photons. When this process occurs spontaneously, with a narrowband pump and under broadband phase matching conditions, time-energy entangled photon pairs are generated, similar to the well known parametric down conversion (PDC) with three waves mixing (TWM) \cite{hong1987measurement,ou1992realization,munro1993violation,joobeur1996coherence,polzik1998subthreshold}. This similarity inspired utilization of spontaneous FWM as a source of entangled photon pairs \cite{edamatsu2004generation,fiorentino2002all,li2005optical,chen2005two,rarity2005photonic,fulconis2007nonclassical,garay_palmett2007photon}. If the pump intensity is increased well beyond the single-pair level, the spontaneous seed can be stimulated, leading to the generation of strong, incoherent, yet highly correlated fields. As was shown previously for PDC, when amplification dominates over the spontaneous seed, this regime of highly amplified spontaneous generation can be well described by a semi-classical model of classical nonlinear amplification applied to a seed gaussian white noise (quantum originated) of order one photon in amplitude per spectral mode \cite{abram1986direct,dayan2004two}.

Both PDC and FWM are governed by similar equations and both, when pumped by a narrowband pump under conditions of broadband phase matching, generate time-energy correlated fields. For PDC the correlation is simple - the signal and idler are complex conjugates $A_s\!\left(\omega\right)\!=\!A_i^*\!\left(\omega_p\!-\!\omega\right)$. This correlation can be measured using two-photon absorption or sum frequency generation (SFG) as a correlation detector. Due to the correlation, the two-photon or SFG spectrum has a sharp peak at the frequency of the original pump due to the broadband constructive interference between correlated signal-idler frequency pairs. This coherent peak responds to coherent manipulations, such as relative delay or pulse shaping, just like an ultrashort pulse, as was observed for TWM with correlation times down to $23fs$  \cite{abram1986direct,dayan2004two,harris2007chirp}.

At first, one expects similar results for correlated fields generated by broadband FWM. Yet, with FWM the phase matching properties depend on an additional parameter- the pump intensity. As we demonstrate here, this introduces surprising coupling effects to the time-energy correlation properties of FWM. In what follows, we describe our experimental measurement of the correlation between the signal and idler using SFG as an ultrafast correlation detector, followed by a theoretical discussion of the observations.

\begin{figure}[h]
\centerline{\epsfig{file=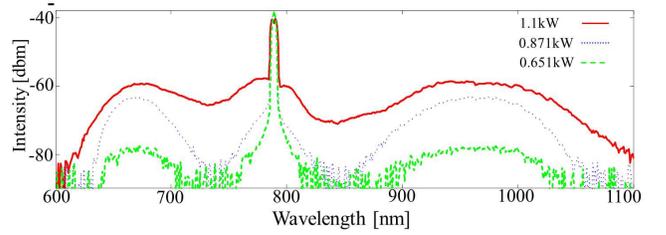, width=1\linewidth}}
\caption{Measured four wave mixing spectra, for three different pump peak power levels - $0.65kW$, $0.87kW$ and $1.1kW$ (pulse duration ~6ps).}
\label{fig:FWMspectrum}
\end{figure}

In the experiment, FWM is generated in a single pass through a short photonic crystal fiber (PCF) (12 cm long polarization maintaining NL-PM-780 from NKT Photonics). The pump is a $6ps$ pulse at $789nm$, near the zero dispersion wavelength of the PCF at $\sim784nm$. The pump pulse is nearly transform limited, narrowband ($<0.3nm$), and is broadened upon propagation by SPM to $1-8nm$ depending on the power level in our experiment. Broad FWM signal and idler sidebands up to tens of mW average power are generated $\sim 100nm$ away from the pump, that are purely initiated by spontaneous emission, as shown in figure \ref{fig:FWMspectrum}, where spectra of pump power dependent FWM are depicted.

In order to measure the correlation of the FWM sidebands, we use SFG as an ultrafast correlation detector. The spectral amplitude of the SFG field is \cite{yariv1976introduction,dayan2004two}:
\begin{equation}
\label{SFG}
\ A_{SFG} \left( \Omega  \right) \propto \int\limits_{\omega} {A_s \left( {\omega '} \right)A_i \left( {\Omega  - \omega '} \right)d\omega '}
\end{equation}
where $A_{s,i}$ are the amplitudes of the signal and idler sidebands entering the SFG crystal. Assuming simple correlation as in PDC, $A\!\left(\omega_0\!+\!\omega\right)\!=\!A^*\!\left(\omega_0\!-\!\omega\right)$, where $\omega_0$  is the center frequency, full constructive interference of all frequency pairs would generate a strong peak at $\Omega\!=\!2\omega_0$: $A_{SFG}\!\left({2\omega _0}\right)\!\propto\!\int{\!\left|{A\left({\omega'}\right)} \right|^2\!d\omega'}$, whereas for $\Omega\!\neq\!2\omega_0$ pairs are uncorrelated leading to a random walk, mostly destructive interference and a low SFG amplitude \cite{dayan2004two}. The correlation properties in frequency and time can be measured by recording the SFG spectrum while scanning a delay imposed between the signal and the idler. With delay $\tau$ the SFG amplitude at $2\omega_0$ is $\int{\!\left|A\!\left(\omega'\right)\right|\!^2\!e^{i\omega'\tau}\!d\omega'}$, and since $\!\left|A\!\left(\omega'\right)\right|^2$ is free of phase, the effect of delay is as if the signal and idler were transform limited pulses of duration $\tau\!\sim\!1/{\Delta}$, where $\Delta$ is the FWM bandwidth \cite{abram1986direct}.

\begin{figure}
\centerline{\epsfig{file=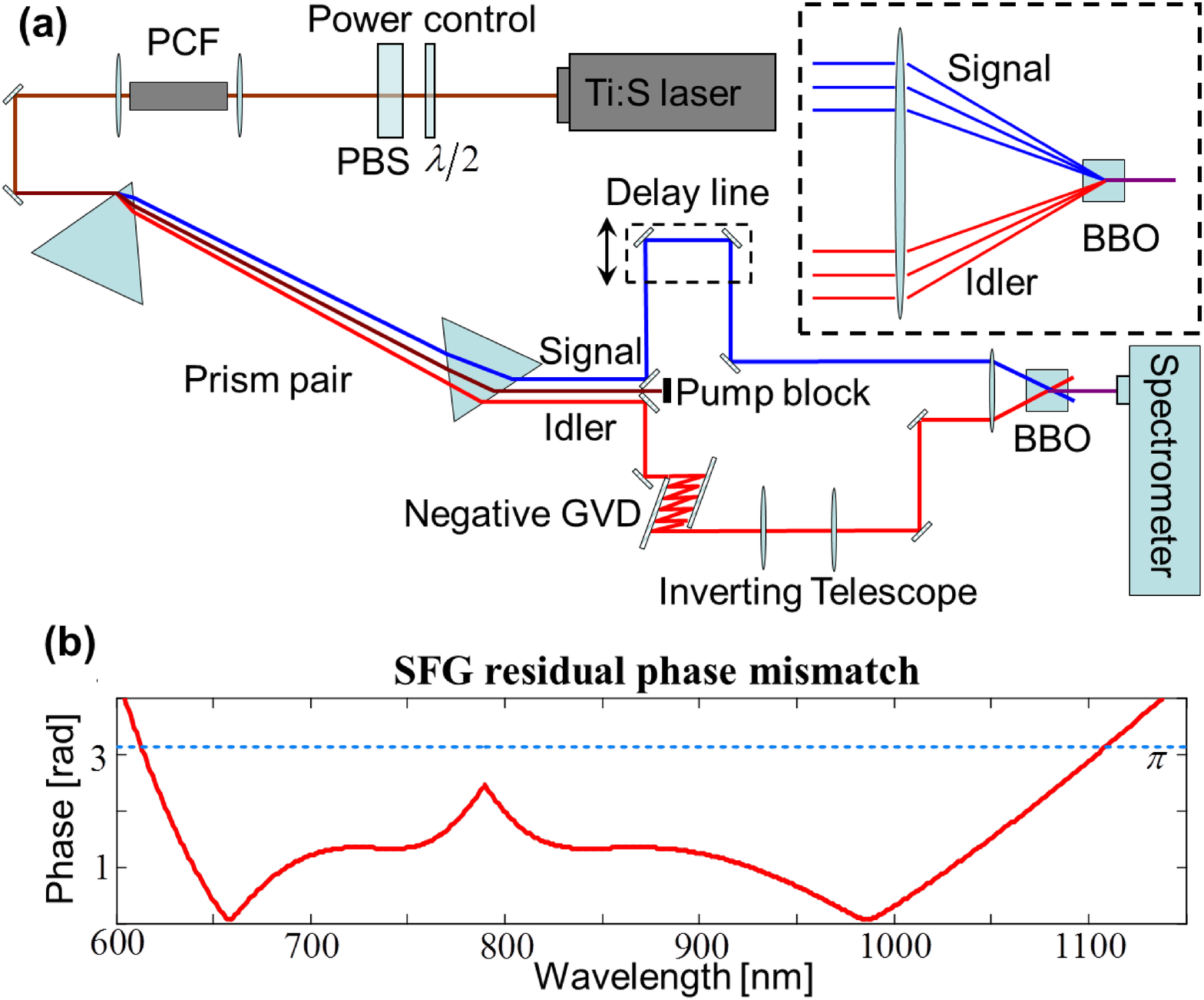, width=1\linewidth}}
\caption{(a) Experimental setup. FWM correlated signal-idler beams are generated by $6ps$ pulses at $789nm$ in a $12cm$ long PCF with zero dispersion at 784nm, and the correlation is measured by SFG in a $200\mu m$ thick BBO crystal, non-collinearly phase matched over the entire spectrum. A prism pair spatially separates the signal, idler and pump, allowing pump blocking, and serves to control dispersion. The signal beam is directed through a computer-controlled delay line with $30nm$ step ($0.1fs$), and the idler beam passes through a magnifying and inverting telescope required to ensure phase matching over the entire signal-idler bandwidth for the non-collinear SFG. The top-right inset is a magnification of the non-collinear SFG configuration. Negative dispersion (in order to compensate for dispersion of lenses, crystals, etc.) is inserted by the prisms and by a set of chirped mirrors in the idler path ($-70fs^2$ per bounce). SFG spectra are measured with a spectrometer coupled to a cooled intensified CCD camera. (b) Calculated residual phase mismatch for SFG into $\Omega=2\omega$ at the BBO crystal.}
\label{fig:experimental_setup}
\end{figure}

The experimental setup (Fig. \ref{fig:experimental_setup}) is composed of two major sub-systems. The first generates correlated FWM; and the second measures this correlation using SFG in a non-linear BBO crystal. Between those two sub-systems auxiliary component are inserted, enabling manipulation of the FWM light, such as spatial separation, pump blocking, spectral filtering, dispersion control, relative delay, etc. (see caption of Fig. \ref{fig:experimental_setup} for details).

As the spectrum is very broad (the signal alone is $\sim 100nm$ wide), SFG can not be phase matched over the entire spectrum in a collinear interaction. We thus resort to non collinear phase matching, and take advantage of the spatial dispersion by the prisms. We arrange the angles of arrival of each frequency component at the SFG crystal such that every signal-idler frequency pair is approximately phase matched for SFG into $2\omega_p$ (see inset of figure \ref{fig:experimental_setup}(a)). Figure \ref{fig:experimental_setup}(b)shows the residual phase mismatch, calculated with realistic parameters. The setup of Fig. \ref{fig:experimental_setup} allows us to compensate group velocity dispersion and scan the relative signal-idler delay, while preserving the non collinear phase matching over the full spectrum.

\begin{figure}
\centerline{\epsfig{file=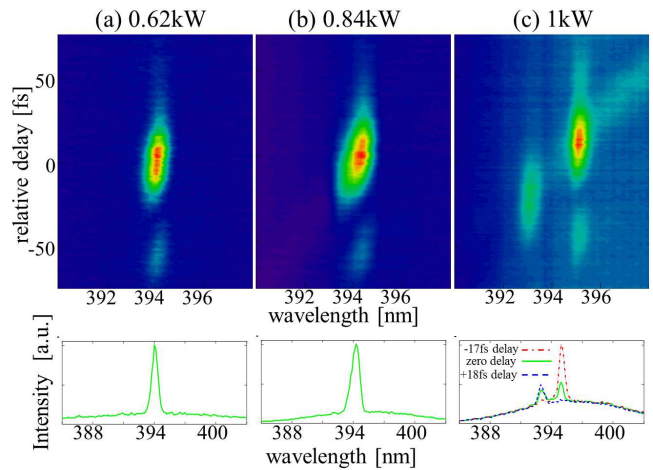, width=1\linewidth}}
\caption{2D maps of SFG spectrum per relative signal-idler delay, and sections from each at zero delay. (a) Low pump pulse intensity ($0.62kW$ peak power). A strong peak at $\Omega=2\omega_p$, with a width of $25fs$, is the signature of two-photon coherence of the signal and idler pairs.  (b) Similar map for a higher pump peak power of $0.84kW$, showing broadening of the coherent peak. (c) High pump peak power ($1kW$) map demonstrating splitting of the coherent peak in both frequency and time. For this intensity the bottom graph shows cros sections at optimum delay for the left peak (blue, dashed line), right peak (red, dash-dotted line) and zero delay (green, full line). The small feature at $-50fs$ delay observed at all intensities may be due to residual high order dispersion.}
\label{fig:colormap}
\end{figure}

We directly measured the time-energy correlation between the signal and the idler by recording spectra of SFG while scanning the delay $\tau$ between the signal and the idler. The results, shown in Fig. \ref{fig:colormap}, exhibit a distinctively different behavior for the low and high pump power regimes. At low power, the SFG demonstrates a single strong peak at $\tau=0$ (Fig. \ref{fig:colormap} (a),(b)), indicating a correlation similar to that observed in PDC. The peak is obtained at $\Omega  = 2\omega _p $ and $\tau=0$ ((Fig. \ref{fig:colormap}(a),(b)- crossections) from the coherent summation of wide spectrum of frequency pairs, whereas at $\Omega  \neq 2\omega _p $ an incoherent background is observed. The temporal width of the coherent correlation peak is $25fs$, close to the expected $20fs$, based on the bandwidth of the FWM sidebands. The $5fs$ discrepancy is mainly attributed to imperfections in the SFG phase matching and dispersion compensation.

As power is increased, the coherent SFG peak splits in both spectrum and relative delay; the intensity of the peaks saturates, while the background intensity continues to increase, leading to a reduction of the contrast between the coherent peaks and the incoherent background (Fig. \ref{fig:colormap} (c)), indicating a more complex correlation. The observed frequency splitting increases with the pump average power and with shortening of the pump pulse, suggesting a dependence on the temporal gradient of the pump pulse intensity. Another important observation is the asymmetry in the SFG spectrum, showing preference for the red side of the spectrum at high pump power.
\begin{figure}
\centerline{\epsfig{file=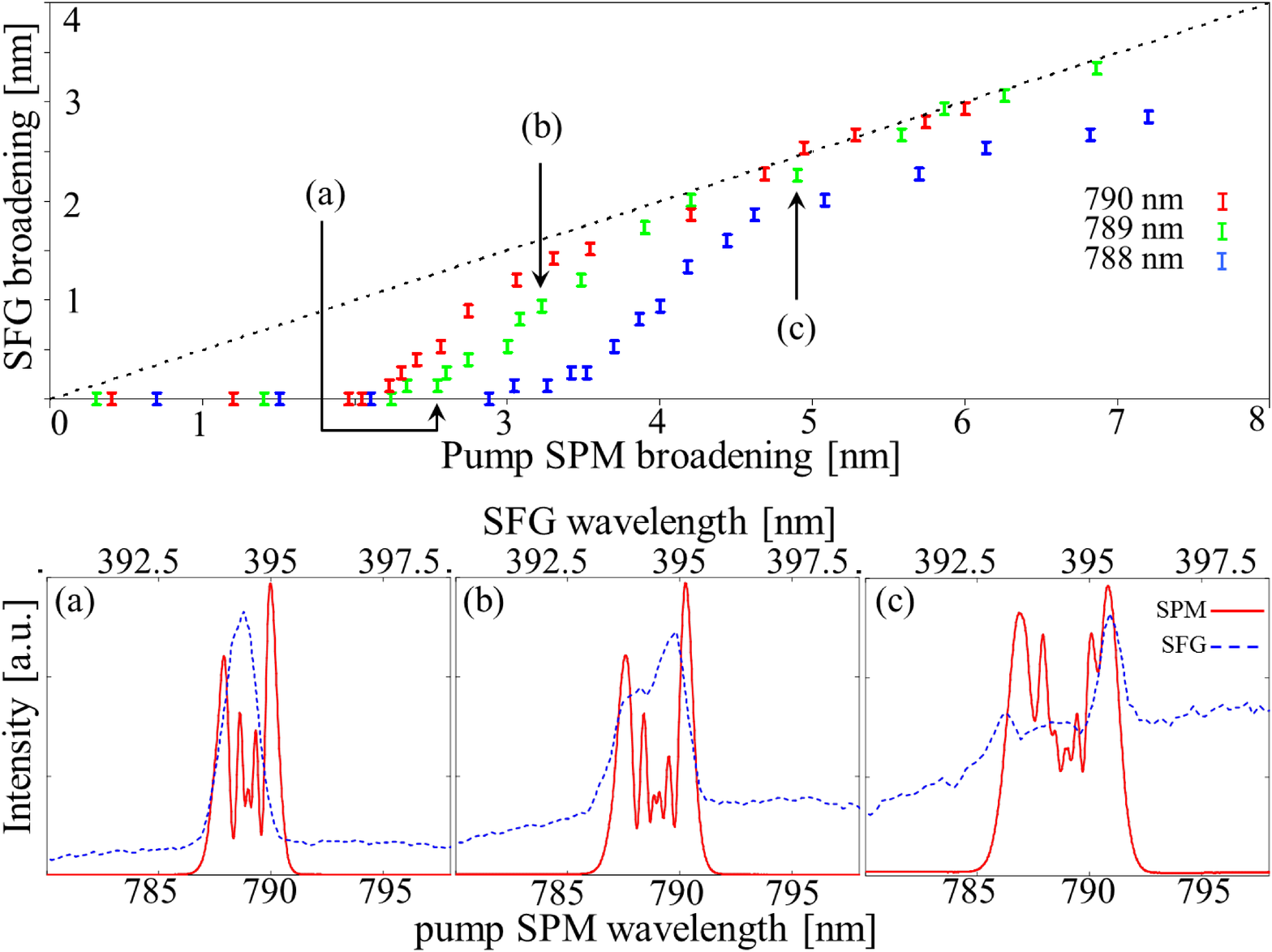, width=1\linewidth}}
\caption{(a) SFG broadening and splitting as a function of the pump broadening. The dashed line marks the identity between the width of SPM broadened pump and the SFG. Clearly identity is followed at high pump splittings (high power) and violated at low pump broadening (low power). For pump wavelengths closer to the zero dispersion of the PCF, the 'transition' from single peaked to split SFG spectrum occurs at increasingly higher pump intensities. The graphs at the bottom show the spectra of the pump after the PCF with the corresponding SFG spectra, measured at the points labeled a, b, c in the top panel.}
\label{fig:SFGvsPump}
\end{figure}

The observed splitting in the spectrum of the SFG peak shows similarity to the spectrum of the pump pulse at the output of the PCF, which is broadened due to SPM. It is therefore reasonable to assume that the splitting of the SFG spectrum is due to cross phase modulation (XPM) of the signal and idler from the SPM broadened pump. To test this assumption, we compared the SFG spectrum to the pump spectrum at the PCF output for various power levels and for several pump wavelengths. As shown in Fig. \ref{fig:SFGvsPump}, one-to-one correspondence is obtained at high powers between the SFG and the pump spectra. At lower power however, the SFG spectrum remains single-peaked also when the pump spectrum is already well split. Moreover, the transition to a split SFG spectrum occurs at a higher pump power (and thus larger pump splitting) as the pump wavelength is varied towards the zero dispersion wavelength of the fiber at $~784nm$. We note that for the signal and idler, variation of the pump wavelength changes considerably the average phase mismatch across the signal-idler spectrum. For the pump however, small variations in wavelength have negligible effect on the pump spectral broadening.

In order to model the SFG spectrum of broadband FWM, let us review the theory of FWM and its correlation properties. Since pump depletion is low and since the pump pulse is narrow and close to the zero dispersion of the medium, dispersion effects on the pump are negligible, and the pump experiences pure self phase modulation as it propagates through the fiber:
\begin{equation}
\ A_p\left(z,t\right)  =A_p\left(z=0,t\right)e^{i\gamma \left| A_p\left(z=0,t\right)\right|^2 z},
\end{equation}
where $A_p$ is the pump amplitude and $\gamma$ is the non-linear coefficient. Thus, for the relatively long pump pulse in our experiment, the time dependent SPM results in a dynamical change of the pump spectral amplitude and phase as it propagates through the PCF.

The time independent equations of FWM for the signal and idler sidebands are \cite{Agrawal2006NL}:
\begin{equation}
\label{FWMeq}
\frac{\partial }{{\partial z}}A_{s,i}  = i\gamma \left( {2\left| {A_p } \right|^2 A_{s,i}  + A_p^2 A_{i,s}^* e^{ - i\left( {\Delta k - 2\gamma \left| {A_p } \right|^2 } \right)z} } \right),
\end{equation}
where the phase mismatch is $\Delta k=2k_p-\left( k_s + k_i \right)$. Rescaling variables for the signal and the idler:
\begin{equation}
\ B_{s,i} = A_{s,i} e^{-i2\gamma \left| A_p\right|^2 z},
\end{equation}
which incorporates XPM from the intense pump, yields:
\begin{equation}
\label{difBsBi}
\frac{\partial }{{\partial z}}B_{s,i}  = i\gamma A_p^2 B_{i,s}^* e^{ - i\Delta \kappa z},
\end{equation}
where $\Delta \kappa$ is the generalized phase mismatch:
\begin{equation}
\label{kappa}
\Delta \kappa = \Delta k + 2\gamma \left| {A_p } \right|^2.
\end{equation}

These equations are similar to the TWM equations, with the important difference that here the field and the phase mismatch are rescaled by cross phase modulation from the intense pump. The solution of eq. \eqref{difBsBi} is:
\begin{equation}
\label{solBsBi}
\ B_{s,i} = \left( a_{s,i} e^{g z} + b_{s,i} e^{-g z} \right) e^{-i \frac{1}{2} \Delta \kappa z},
\end{equation}
where the gain is:
\begin{equation}
\label{gain}
\ g = \sqrt{\gamma^2 \left| A_p \right|^4 - \frac{1}{4} \Delta \kappa^2 } = \sqrt{-\Delta k\gamma \left| A_p \right|^2 - \frac{1}{4} \Delta  k^2}.
\end{equation}

In spite of the similarity to TWM \cite{yariv1976introduction}, it is clear that considerable differences exit, such as the dependence of the generalized phase mismatch on the pump intensity. In addition, for TWM, gain occurs only near $\Delta k\!=\!0$, whereas for FWM gain exists also for very large generalized phase mismatch ($\Delta\kappa l\!\gg\!2\pi$, where $l$ is the medium length). As evident seen from eq.\eqref{gain}, for $\Delta k\!<\!0$ the gain increases with pump power, regardless of the resulting $\Delta \kappa$. At high power ($\gamma\left| A_p \right|^2\!\gg\!\left|\Delta k\right|$), the gain is proportional to $\sqrt{\Delta k}$ ($g\!\approx\!\sqrt{\!-\!\Delta k\gamma\!\left| A_p \right|^2}$). However, the maximum gain for a given pump intensity $g\!=\!\gamma\left| A_p \right|^2$ is obtained for ideal generalized phase mismatch, $\Delta \kappa\!=\!0$ (which occurs only when $\Delta k\!=\!-\!2\gamma\left| A_p \right|^2\!<\!0$).

The dependence of the gain on $\Delta k$ can explain the asymmetry in the SFG spectrum with preference towards the red. Since the bare phase mismatch $\Delta k$ at the signal and idler wavelengths depends quadratically on the deviation of the pump from the zero dispersion wavelength, pump wavelengths further to the red will have a larger $\Delta k$, and hence larger gain. Consequently, the FWM generated by the red side of the pump spectrum will be more pronounce than FWM generated by the blue side, leading to the observed asymmetry in the SFG.

The temporal splitting of the correlation peak is explained by the slight difference in group velocity between signal-idler pairs that sum up to each of the SFG spectral lobes. Specifically, the relative group delay $\tau\!=\!\left(\beta_s\!-\!\beta_i\right)l$ (where $\beta_{s,i}\!=\!{\partial k_{s,i}}/{{\partial \omega}}$ are the inverse group velocities of the signal and the idler fields compared to the pump) depends on the center frequency shift $\delta$, leading to a slightly different time delay for each coherent peak. While these frequency and time shifts are minute compared to the bandwidth and duration of the sidebands, they are readily measured from the SFG spectrum.

The spectral splitting of the SFG spectrum is well understood as a result of XPM from the intense pump to the signal and idler fields. At this point however, we do not fully understand the threshold-like behavior of the SFG splitting (Fig. \ref{fig:SFGvsPump}), which occurs only when the pump broadening surpasses a certain threshold. The dependence of the threshold position on pump wavelength hints though, that this effect is also related to the value of the gain. Indeed, as the pump wavelength approaches the zero dispersion point, the parametric gain per given pump power is reduced, and the threshold for SFG splitting is pushed to higher pump power.

Our experiments on the coherence properties of the FWM amplifier and theoretical model highlight the striking difference between stimulated and spontaneously seeded FWM. Since FWM amplification is the key process for frequency comb generation in micro-cavities, understanding its coherence properties is crucial for analysis of the generated comb. Let us try to elaborate on the relation of our work here to the emergence of broadband coherence in FWM micro-cavity oscillators: Frequency combs are first-order (single photon) broadband coherent fields with a definite phase relation between all equi-spaced teeth. Spontaneous FWM however generates two-mode coherence, but not first order coherence as we showed. The question then becomes - how can broadband first order coherence emerge in FWM comb oscillators, which rely on spontaneous FWM as a seed? Clearly, the first mode pair is generated from amplified spontaneous FWM with a random phase $\phi$ between the signal and the pump. However, due to the two-mode coherence, the pump-idler phase will also be $\phi$. If additional pairs would be also generated spontaneously, their phase will be unrelated to $\phi$ and single-photon coherence will not emerge. If however, the same phase $\phi$ is heralded by stimulation to all adjacent teeth in the oscillation spectrum, the stable spectral phase needed for comb may form. This would require a "cascade process", which can flow as follows: Once the first signal-idler pair is well amplified, the resulting three modes can generate more modes through stimulated non-degenerate FWM. Specifically, the "old" pump+signal can act as new pump fields for the next generation which will produce from the "old" idler a new signal mode on the next tooth, with the same near-neighbour phase $\phi$, and so on. In this way only one mode pair is generated from the spontaneous seed (randomly selecting $\phi$), but all additional comb teeth are generated by Stimulated FWM, distributing the same phase $\phi$ across the entire comb. This process would be hampered of course, if more than one initial mode pair is spontaneously generated. The different pump-signal-idler trios may then yield different sub-combs through the cascaded process mentioned above, but these sub-combs will not be mutually coherent, as indeed was recently observed \cite{okawachi2011octave, ferdous2011spectral,johnson2012chip,ferdous2012probing}.

In conclusion, broadband FWM pumped by $ps$ pulses demonstrates unique two-photon coherence, different from its well known counterpart of PDC. Using SFG as an ultrafast two-photon correlation detector, we demonstrated time-energy coupling effects unique to FWM, due to XPM of the FWM sidebands from the intense pump. Our findings may serve as a first step towards understanding the observed coherence in FWM comb oscillators, where stimulation and spontaneous seeding are intertwined.

This research was supported by the Marie Curie International Reintegration Grant (IRG) under EU-FP7.

\bibliographystyle{apsrev}

\end{document}